\documentclass[prl,twocolumn]{revtex4-1}
\usepackage{amsfonts}
\usepackage[T1]{fontenc}

\usepackage{amsmath,amsbsy,amssymb,graphicx}
\usepackage{times}
\usepackage{color}
\usepackage{upgreek}

\let\mathbf=\boldsymbol

\begin{document}

%\title{{\Large Perimeter Excitation Modes of a Magnetic Droplet Soliton}}
\title{{\Large Parametric Auto-Excitation of Magnetic Droplet Soliton Perimeter Modes}}

\author{D. Xiao$^{1,7}$}
\author{V. Tiberkevich$^{2}$}
\author{Y. H. Liu$^{3,8}$}
\author{Y. W. Liu$^{1}$}
\email[Corresponding author:~]{yaowen@tongji.edu.cn}
\author{S. M. Mohseni$^{4,6}$}
\author{S. Chung$^{5,6}$}
\author{M. Ahlberg$^{5}$}
\author{A. N. Slavin$^{2}$}
\author{J. \AA kerman$^{5,6}$}
\email[Corresponding author:~]{johan.akerman@physics.gu.se}
\author{Yan Zhou$^{7}$}
\email[Corresponding author:~]{zhouyan@cuhk.edu.cn}
\affiliation{$^{1}$Shanghai Key Laboratory of Special Artificial Microstructure Materials and Technology, School of Physical Science and Engineering, Tongji University, Shanghai 200092, China}
\affiliation{$^{2}$Department of Physics, Oakland University, Rochester, Michigan 48309, USA}
\affiliation{$^{3}$
Theoretische Physik, ETH Zurich, 8093 Zurich, Switzerland}
\affiliation{$^{4}$
Department of Physics, Shahid Beheshti University, Tehran 19839, Iran}
\affiliation{$^{5}$
Department of Physics, University of Gothenburg, 412 96 Gothenburg, Sweden}
\affiliation{$^{6}$
Materials Physics, School of Information and Communication Technology, KTH Royal Institute of Technology, Electrum 229, 164 40 Kista, Sweden}
\affiliation{$^{7}$School of Science and Engineering, Chinese University of Hong Kong, Shenzhen, 518172, China.}

\begin{abstract}
Recent experiments performed in current-driven nano-contacts with strong perpendicular anisotropy have shown that spin-transfer torque can drive self-localized spin waves \cite{Rippard2010,Mohseni2011} that above a certain threshold intensity can condense into a highly nonlinear magnetodynamic and nano-sized state known as a magnetic droplet soliton  \cite{Mohseni2013}. Here we demonstrate analytically, numerically, and experimentally that at sufficiently large driving currents, and for a spin polarization that is tilted away from the film normal, the circular droplet soliton can become unstable to periodic excitations of its perimeter. We furthermore show that these perimeter excitation modes (PEMs) are parametrically excited when the fundamental droplet soliton precession frequency is close to twice the frequency of one or more of the PEMs. As a consequence, for increasing applied fields, progressively higher PEMs can be excited. %, including linear combinations of neighboring PEMs.
Quantitative agreement with experiment confirms this picture.\end{abstract}

\maketitle
\noindent\textbf{\uppercase\expandafter{\romannumeral1}. INTRODUCTION}

The theoretical, numerical, and experimental study of non-trivial nanoscale spin structures, such as domain walls, vortices, anti-vortices, skyrmions, merons, magnetic bubbles, and spin wave "bullets", has recently attracted a lot of interest \cite{Slavin2005,Braun2012,Mohseni2013,Zhou2014b,Ruotolo2009,Phatak2012,Ezawa2011,Gliga2013,Muehlbauer2009,Yu2010,Nagaosa2013,Fert2013,Zhang2016,Liu2015b,Lin2013a,Zhou2014a,Jiang2015,Hoffmann2015,Boulle2016,Woo2016,Moreau-LuchaireC2016}. The control and manipulation of these static or dynamic spin structures are important both from the point of view of fundamental understanding of nanoscale magnetism, and the possible applications of some of them as information carriers or microwave signal generators in future spin-based devices \cite{Dumas2014,Zhou2014b}. The recent experimental observation of magnetic droplet solitons \cite{Mohseni2013,Mohseni2014,Macia2014,Chung2014jap,Lendinez2015,Chung2015ltp,Chung2016a,Backes2015}  (droplets from hereon) in nanocontact (NC) spin torque oscillators based on a material with large perpendicular magnetic anisotropy (PMA) \cite{Rippard2010,Mohseni2011}, adds a new member to this family of distinct and useful nanoscale magnetic objects. In contrast to the bullet \cite{Slavin2005}, which is formed in in-plane \cite{Rippard2004,Demidov2010ntm}, or close to in-plane \cite{Bonetti2010,Bonetti2012prb,Dumas2013prl}, magnetized films, %with a negative nonlinear frequency shift,
the droplet appears in out-of-plane magnetized films with strong perpendicular anisotropy \cite{Hoefer2010}. The existence of its conservative sibling, the magnon drop, was predicted in lossless materials with PMA \cite{Ivanov1977,KOSEVICH1990}, and the droplet, driven by spin transfer torque (STT) underneath a NC, inherits most of its dynamics \cite{Hoefer2010,Hoefer2012a,Hoefer2012, Iacocca2014}.

The droplet can be described as a partially inverted magnetic domain with all its spins precessing in phase at a frequency lying inside the spin-wave gap of the film. In addition to the fundamental uniform precession, droplets can exhibit internal dynamics observed experimentally as additional microwave signals appearing at lower frequencies \cite{Mohseni2013,Lendinez2015,Chung2016a}. In this work, we consider one particular type of such internal dynamics - the perimeter excitation modes (PEMs) associated with periodic spatial deformations of the droplet perimeter. We begin by analytically deriving the PEM eigen-frequencies and spatial profiles for a conservative magnon drop (hence ignoring damping and STT) and study their dynamics numerically. We then show that PEMs can also be spontaneously excited in STT driven droplets and identify the excitation mechanism as \textit{parametric}, i.e. a PEM is predominantly excited when the fundamental droplet frequency is twice that of the PEM. Finally, we compare our model and simulations with experimental results and find an excellent agreement.\\

\noindent\textbf{\uppercase\expandafter{\romannumeral2}. PERIMETER MODE DERIVATION}

The Landau-Lifshitz-Gilbert-Slonczewski (LLGS) equation describing dissipative and current-driven magnetization dynamics in a thin ferromagnetic film with perpendicular anisotropy has the form \cite{Slonczewski1999}:
\begin{equation} \label{llgs}
\begin{aligned}
d\mathbf{m}/dt=&-\mathbf{m}\times\mathbf{h}_{eff}-\alpha\mathbf{m}\times(\mathbf{m}\times\mathbf{h}_{eff})\\
&+\sigma\mathbf{m}\times(\mathbf{m}\times\mathbf{m}_{f})
\end{aligned}
\end{equation}
where $\mathbf{m} = \mathbf{m}(t, \mathbf{r}) = \mathbf{M}$/M$_s$ is the unit magnetization vector and M$_s$ the saturation magnetization. The first term on the right-hand side describes the magnetization precession in an effective field $\mathbf{h}_{eff}$, which includes the contributions from the inhomogeneous exchange, easy $z$-axis anisotropy characterized by the magnetic induction $B_A$ and an external perpendicular bias magnetic field $H$. The second term is the Gilbert damping with $\alpha$ being the dimensionless Gilbert damping parameter. The third term describes the STT, where $\sigma$ is the parameter characterizing the magnitude of the driving spin-polarized current, and $\mathbf{m}_{f}$ is a unit vector defining the direction of the spin polarization in the driving current. This direction coincides with the magnetization direction in the "fixed" magnetic layer of the NC spin torque oscillator trilayer.

For our initial analytical treatment, we ignore both damping and STT and derive the perimeter dynamics of a conservative magnon drop with the fundamental precession frequency $\omega$ given by the following approximate expression \cite{Ivanov1977,Hoefer2010,Ivanov1989}:
\begin{align}
\omega_0(H)=\omega_H+\dfrac{\lambda_{ex}\omega_M}{\rho_0}
\label{omega0s}
\end{align}
where $\omega_H=\gamma H$, $\omega_M=\gamma \mu_0 M_s$, $H$ is the magnitude of the perpendicular bias magnetic field, $\rho_0$ is the effective radius of the drop, $\gamma$ is the modulus of the gyromagnetic ratio for electron spin, $\mu_0$ is the magnetic permeability of vacuum, $\lambda_{ex} =\sqrt{A_{ex}/K}$   is the exchange length, $A_{ex}$ is the exchange constant.

In the cylindrical coordinate system ($\rho, \varphi$) the spatial profile of the drop can be written as \cite{Ivanov1977,Hoefer2010}
\begin{equation} \label{m0}
\begin{aligned}
\mathbf{m}_0(\rho,\varphi,t)=&[\sin\theta(\rho)\cos\Phi(t),\sin\theta(\rho)\sin\Phi(t),\\
&\cos\theta(\rho)],
\end{aligned}
\end{equation}
where the phase $\Phi=\arctan (m_y/m_x)$ is the same at all the points in space, and evolves linearly in time:
\begin{align}
\Phi(\rho,\varphi,t)=\omega_0t+\Phi_0.
\label{Phi}
\end{align}
Here $\theta(\rho)$ gives the shape of the drop, which satisfies $\theta(0) < \pi$ and $\theta(\infty) = 0$. The in-plane component of the magnetization in the drop points in the same direction in the whole plane, and precesses at the fundamental frequency $\omega_0$ (see Fig. 1). In the fundamental dynamics, the the \textit{z}-component of the magnetization does not change with time at any point. For an almost fully reversed droplet, the magnetization both in the center and far away from the drop is also practically static, as is clear from Eq.(\ref{m0}) and Eq.(\ref{Phi}). The most interesting dynamical region is hence the perimeter, which we define as a line of zero perpendicular magnetization ($\theta = \pi/2$), and a surrounding region with a width approximately given by the exchange length.

A convenient way to describe the PEMs is to introduce the dimensionless dilatation factor
\begin{align}
	a(\varphi, t) = \rho(\varphi, t)/\rho_0 - 1
	\label{defa}
\,,\end{align}
and expand $a(\varphi, t)$ into a Fourier series,
\begin{align}
	a(\varphi,t)=\sum_{n=-\infty}^{+\infty}a_n(t)\exp(in\varphi).
\label{rhop}
\end{align}
%we found no transfer of weight from the original PEM into any other PEM for all modes investigated here.
Each component $a_n(t)$ varies with time as
\begin{align}
a_n(t)\propto\cos(\Omega_nt)\exp(-\Gamma_nt)
\label{tilderhop}
\end{align}
where $\Omega_n$ is the PEM eigen-frequency and $\Gamma_n$ is the mode dissipation parameter.

In a general case, the position of the perimeter can be parametrized by a certain radial function $\rho_p(\varphi, t) = [1+a(\varphi,t)]\rho_0$, while the phase of the in-plane magnetization components at the perimeter can be parametrized by the phase $\Phi_p(\varphi, t)$. A perfect drop has a circular static perimeter with vanishing dilatation factor, $a(\varphi,t) = 0$, and a uniform phase Eq.(\ref{Phi}). To describe the dynamics of a slightly deformed drop we represent the phase $\Phi(\varphi,t)$ as
%\begin{subequations}  \label{rhops}
\begin{align}
%\rho_p(\varphi,t)&=[1+a(\varphi,t)]\rho_0\\
\Phi(\varphi,t)=\omega_0t+\psi(\varphi,t)
\end{align}
%\end{subequations}
and assume that both the dilatation factor $a(\varphi, t)$ and phase deformation $\psi(\varphi,t)$ are small quantities.
Using the method developed in Ref. [\onlinecite{Clarke2008,Makhfudz2012}] we find the small-amplitude equations for $a(\varphi, t)$ and $\psi(\varphi, t)$:
\begin{subequations} \label{coupledEq}
\begin{align}
\partial a(\varphi,t)/\partial t&=\Omega\partial^2\psi/\partial\varphi^2,\\
\partial\psi/\partial t&=-\Omega(a+\partial^2a/\partial\varphi^2),
\end{align}
\end{subequations}
where

\begin{align}
\Omega=\gamma(2K/M_s-\mu_0M_s)\cdot(\lambda_{ex}/\rho_0)^2
\label{Omegas}
\end{align}
is the characteristic frequency of the perimeter excitation modes (PEMs).

The eigen-solutions of Eq.(\ref{coupledEq}) describe possible low-amplitude oscillations of the perimeter. There is an infinite number of perimeter oscillations which can be enumerated by an integer $n$ describing the angular dependence of the amplitude $a(\varphi, t)$:
\begin{align}
a_n(\varphi,t)=A_n\exp(-i\Omega_nt+in\varphi)
\label{anvarphit}
\end{align}
where the PEM eigen-frequency $\Omega_n$ is given by:
\begin{align}
\Omega_n=n\sqrt{n^2-1}\Omega.
\label{Omegans}
\end{align}
\\
From our derivation of the PEMs of conservative drops, we can draw several important conclusions. First, the PEMs with positive and negative indices $\pm n$ are degenerate, %which means that
\emph{i.e.} one may expect to observe PEMs in the form of standing waves. %($\propto \cos(n\varphi)$).
In addition, the frequencies for $n=0$ and $n=\pm1$ vanish, which means that these PEMs correspond to unstable deformations of the perimeter. The circularly-symmetric mode $n=0$ describes the expansion/contraction of the drop as a whole. Since the conservative drop radius $\rho_0$ can take any value, there is no restoring force for such uniform dilatation, and the mode frequency vanishes. Note, however, that the dissipative droplet radius is always of the order of the NC radius \cite{Hoefer2010}, set by the balance between damping and the locally supplied STT [both ignored in Eq. (\ref{anvarphit})], and will experience a restoring force; the $n = 0$ mode for the droplet hence has a non-vanishing frequency (see below). Similarly, the two modes $n = \pm1$ describe the lateral shift of the drop without changes in its profile and are again marginally stable in a conservative case. Since the droplet, on the other hand, is confined to the NC region, it again experiences a restoring force, which leads to a finite frequency also for the two $n = \pm1$ modes.\\

\noindent\textbf{\uppercase\expandafter{\romannumeral3}. MICROMAGNETIC SIMULATIONS}

To check these analytical predictions we carried out micromagnetic simulations of magnon drops, where a PEM perturbation of the perimeter was introduced as an initial condition. At selected time steps, we recorded the configuration of the $z$-component of the magnetization and used numerical interpolation to find the perimeter function $\rho_p$($\varphi$,$t$), with $\varphi$ being the azimuthal angle. The radius of the unperturbered drop was $\rho_0$ = 58 nm, which in the case of a Co/Ni multilayer free layer ($M_s$ = 300 kA/m \footnote{The low value for $M_s$ was chosen to avoid substantial overlap of the different PEMs and corresponds to Ni rich Co/Ni multilayers.}) corresponds to a characteristic PEM frequency of $\Omega = 9.4/2\pi$ GHz. The magnetic field was set to zero and a magnetic damping of $\alpha$ = 0.01 was added. Micromagnetic simulations are performed using Mumax3 \cite{Vansteenkiste2014}. We consider a 256$\times$256 square lattice with a unit cell size of 2 nm$\times$2 nm$\times$2 nm. In Fig. 1-3, the  material parameters adopted in the simulations for Co/Ni free layer based NC-STO are as follows: thickness = 2 nm, NC radius $R_c$ = 50 nm, exchange stiffness $A_{ex}$ = 30 pJ/m, and fixed layer spin polarization ratio $P$ = 0.5. %, with the spin-polarization vector $\mathbf{m}_f$ forming an angle of $\theta_p$=45 degree with respect to the \textit{z} axis.
In Fig. 4, the material parameters used in our simulations are the similar to those of the experiment %of NC-STO based on Co/Cu/Co-[Ni/Co]$_{\times4}$ orthogonal spin-valve
\cite{Mohseni2013, Iacocca2014}: $R_c$ = 60 nm, $A_{ex}$ = 30 pJ/m, $K$ = 447 kJ/m$^3$, $P$ = 0.5 and $M_s$ = 716 kA/m. All simulations are performed at zero temperature. The current-induced Oersted field is included %in the simulations
assuming a current flow %path as
through an infinite cylinder.

\begin{figure}[h!]
\includegraphics[width=0.48\textwidth]{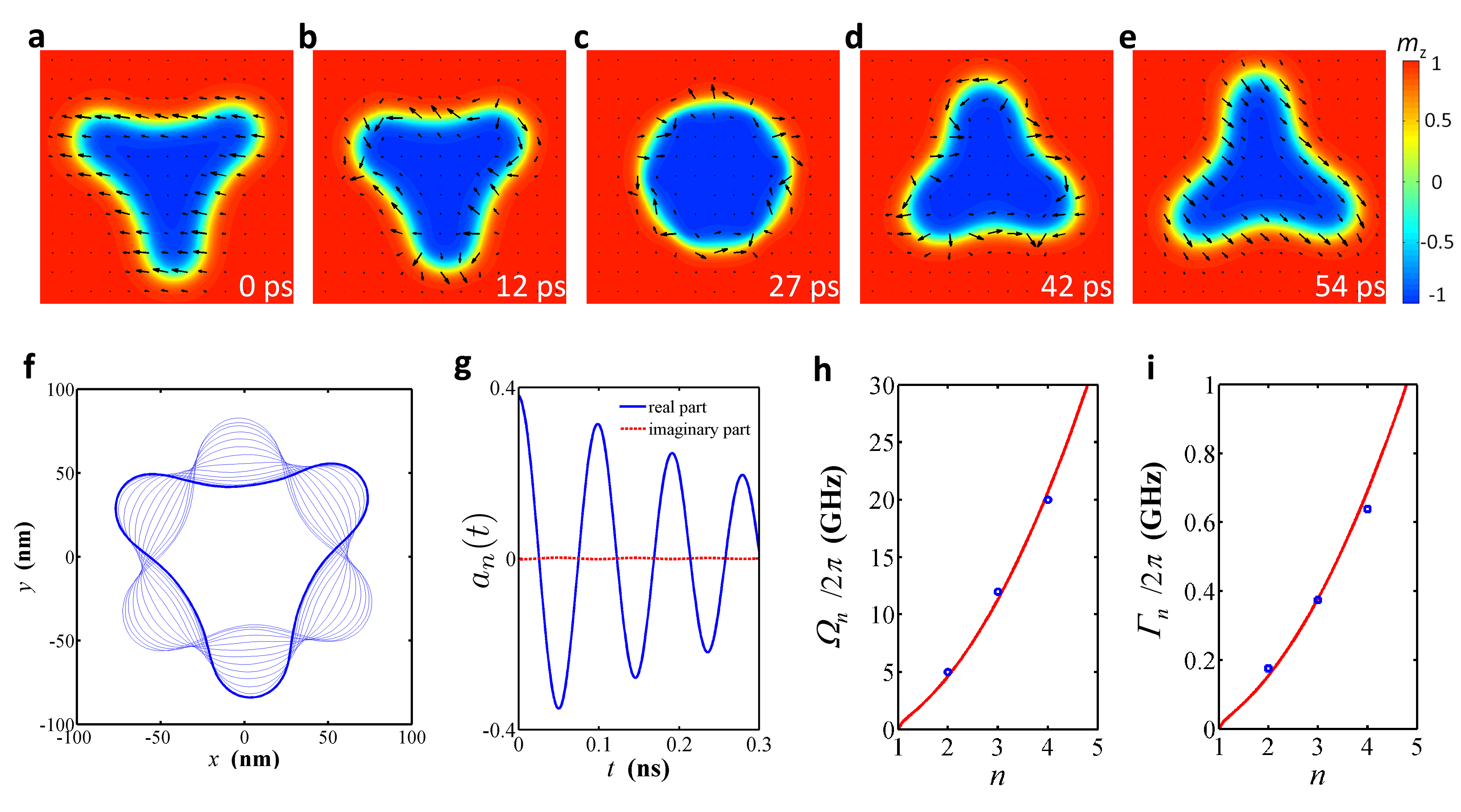}
\caption{{Perimeter excitation modes (PEM).} (a-e) Two-dimensional snapshots of the magnon drop PEM mode $n$ = 3 during one half-period of oscillations. The color code shows the magnitude of the out-of-plane $m_z$ component of the magnetization, with the arrows indicating the in-plane component. (f) Overlayed snapshots from the time evolution of the shape of the drop.
%(g) of the numerically simulated MDS perimeter function $\rho_p (\varphi, t)$ for $n$ = 3.
(g) The numerically simulated PEM amplitude $a_n(t)$ for $n=3$.
Dots in the frames (h) and (i) show the numerically calculated magnitudes of the PEM frequencies $\Omega_n$ and damping rates $\Gamma_n$ as functions of the PEM index $n$. Solid lines in these frames are showing the results of analytical calculations of the same quantities using Eq. (\ref{Omegans}) and Eq. (\ref{Gamman}), respectively, for $\Omega/2\pi$ = 1.5 GHz [see Eq. (\ref{Omegas})] and $\alpha_p = 0.031$.}
\label{Fig:PEMmodes}
\end{figure}

Snapshots of the magnetization for a PEM with mode index $n$ = 3, amplitude $A_3  = 0.4$, and uniform phase, are shown in Fig. \ref{Fig:PEMmodes} (a)-(e) during half a period of perimeter oscillation; red (blue) color corresponds to a positive (negative) $z$-component of the magnetization. The initial state first evolves through a rapid increase in the non-uniformity of the perimeter phase, nucleated at the three points of maximum curvature, where the in-plane component of the magnetization rapidly changes sign.
%This is followed by a more gradual reduction of the dilatation of the radius $a$($\varphi$, $t$), which makes the perimeter essentially regain its non-perturbed circular shape after a quarter period ($t$ = 27 ps), however with a maximum degree of phase inhomogeneity.
This is followed by a gradual reduction of the distortions of the perimeter function $\rho_p(\varphi, t)$, which makes the perimeter essentially regain its non-perturbed circular shape after a quarter period ($t$ = 27 ps), however with a maximum degree of phase inhomogeneity.
Past this point, the perimeter redevelops radial deformations while the phase gradient decreases. After half a period ($t$ = 54 ps) the phase again becomes uniform, and the radial dilatation again reaches a maximum, now with opposite sign of the in-plane magnetization compared to the initial condition. This coupled evolution of radius and phase deformations proves their connection described analytically by Eq. (\ref{coupledEq}).

%The fact that the perturbed drop does not noticeably change its shape between Fig.\ref{Fig:PEMmodes} (a) and Fig.\ref{Fig:PEMmodes} (e) excludes any nonlinear mixing between different PEMs.

The time evolution of the real part \footnote{The imaginary part is always zero, which means the  angular velocity of the collective rotation of the deformed perimeter is 0}  of $a_n(t)$ (Fig. \ref{Fig:PEMmodes}(g)) is adequately described by (\ref{tilderhop}) with $\Omega_3$ = 79.8$/2\pi$ GHz, and $\Gamma_3$ = 2.35$/2\pi$ GHz. Fig. \ref{Fig:PEMmodes}(h) shows $\Omega_n$ for PEMs with indices $n$ = 2-4, together with a solid line calculated analytically using Eq. (\ref{Omegans}) with $\Omega/2\pi$ = 1.5 GHz, which corresponds to the parameters of our numerical simulations. It is clear that the simple analytic expression describes the numerically calculated PEM frequencies very well. Fig. \ref{Fig:PEMmodes}(i) shows the corresponding $\Gamma_n$ with a solid line calculated analytically assuming that the PEM damping is proportional to the PEM frequency  $\Omega_n$,
\begin{align}
\Gamma_n=\alpha_p\Omega_n
\label{Gamman}
\end{align}
with an "effective Gilbert damping" constant equal to $\alpha_p$ = 0.031.

\begin{figure}[h!]
\centering
\includegraphics[width=0.48\textwidth]{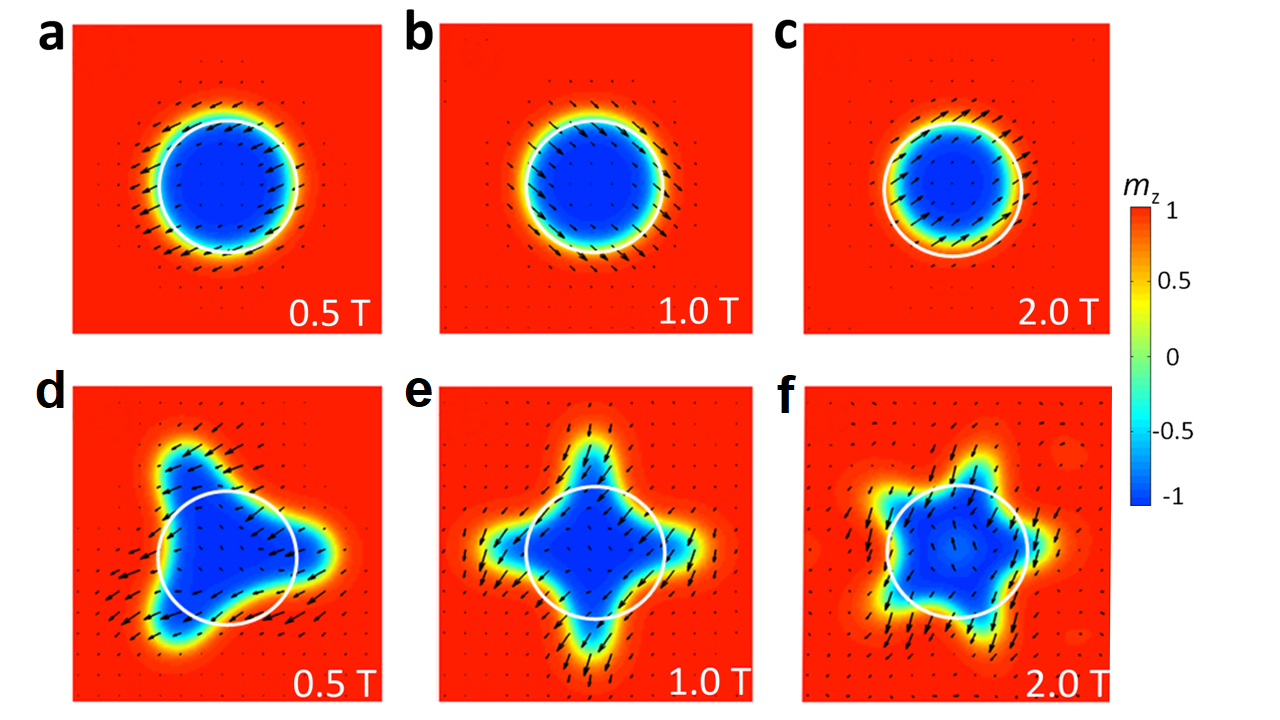}
\caption{\noindent{{Spin transfer torque driven magnetic droplet solitons and the perimeter excitation modes.} (a) - (c) Snapshots of the free layer magnetization in a micromagnetically simulated NC spin torque oscillator at a low current density ($j$ = 0.2$\times10^{12}$ A/m$^2$). (d) - (f): Same simulation as in panels (a) - (c), however at a much higher current density ($j$=2.5$\times$10$^{12}$ A/m$^2$). High amplitude PEM excitations are observed at progressively higher mode numbers when the applied field is increased. The color indicates the perpendicular component $m_z$ with red (positive $m_z$) corresponding to spins aligned with the applied field, and blue (negative $m_z$) to spins pointing away from the applied field. The black arrows provide the corresponding in-plane direction and magnitude of the magnetization. The white circle outlines the NC.}}
\label{Fig:newfigure2}
\end{figure}

\begin{figure}[h]
\centering
\includegraphics[width=0.48\textwidth]{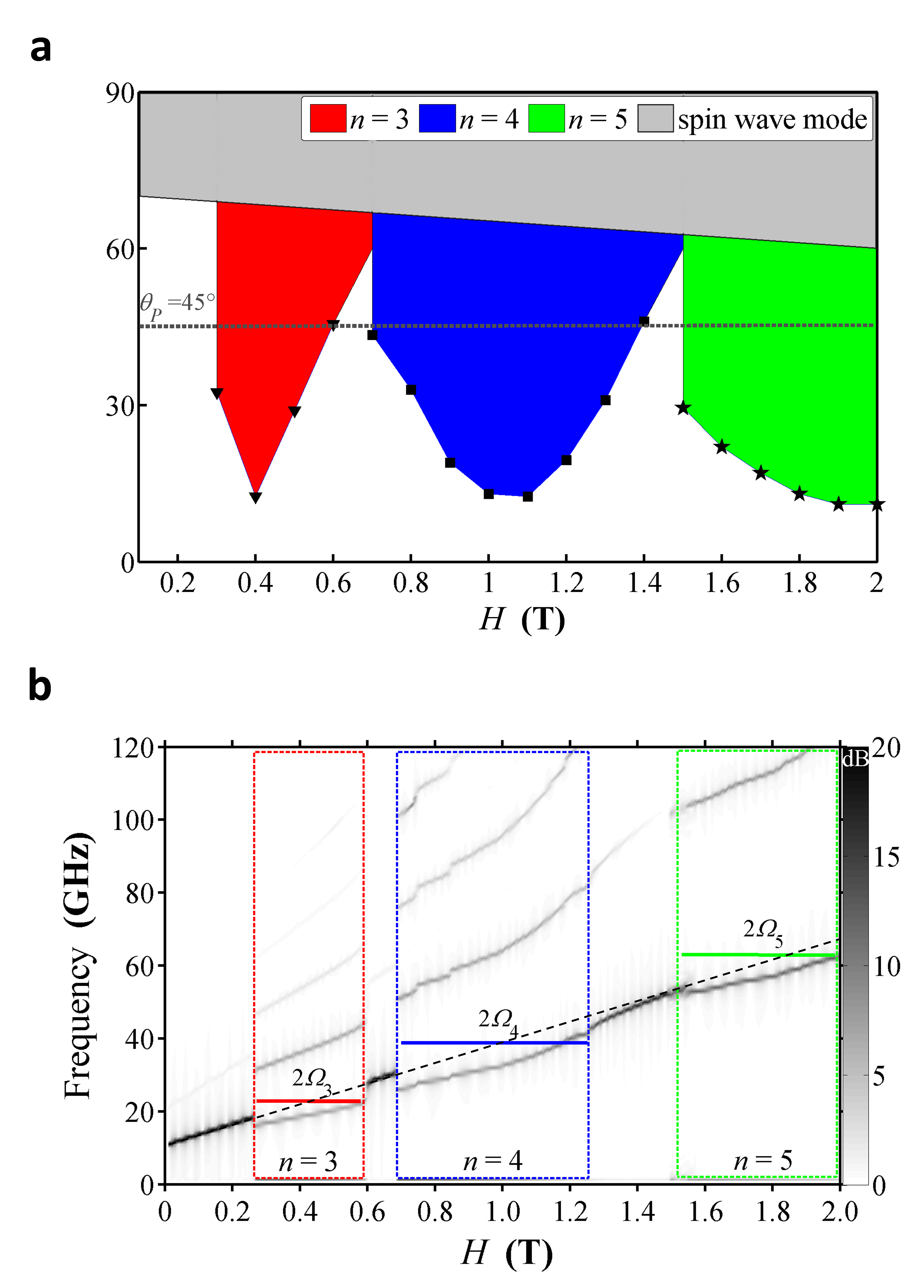}
\caption{\noindent{Phase diagram of droplet PEMs parametrically excited by the uniform droplet precession.} The current density is $j$=2.5$\times$10$^{12}$ A/m$^2$. High amplitude PEM excitations are observed at progressively higher mode numbers when the applied field is increased.}
\label{Fig:phasediagram}
\end{figure}

As seen in Fig. \ref{Fig:newfigure2}(a)-(c) the circular droplet remains stable for all values of the bias magnetic field when the current density is small ($j=0.2\times10^{12}$ A/m$^2$). However, if the current density is increased an order of magnitude, the droplet perimeter loses its stability through the excitation of one of the PEMs %(Fig.\ref{Fig:PEMhigherJ} (a)-(c)),
(Fig. \ref{Fig:newfigure2} (d)-(e)), and when the magnetic field is increased, PEMs having progressively higher indices $n$ are excited. %We summarize our simulations as a phase diagram
In Fig. \ref{Fig:phasediagram}(a) we vary both the applied field strength and the spin polarization angle $\theta_p$  of the NC current (i.e. the tilt angle of the fixed layer). It is clear that PEMs only appear above a certain threshold  $\theta_p^{th}$ and that the field intervals where PEMs are observed increase with increasing $\theta_p$.

In Fig. \ref{Fig:phasediagram}(b) we show the Fourier spectrum of the droplet precession as a function of magnetic field for $\theta_p = \pi/4$. The dashed line indicates the approximate analytical expression for the fundamental precession [see Eq. (\ref{omega0s})]. The PEM regions with $n$ = 3, 4, 5 are shown by colored boxes and within each box we have highlighted 2$\Omega_n$, i.e. \textit{twice} the PEM frequency for each mode. We conclude that PEMs are only excited when the droplet precession frequency is close to twice that of the PEM, which strongly indicates that the latter is parametrically excited by the former. This parametric process has the energy conservation law \cite{Urazhdin2010,lvov1994}:
\begin{align}
\omega_0(H)=\Omega_n+\Omega_{-n}=2\Omega_n
\label{omega0}
\end{align}
and at the threshold value of the polarization angle $\theta_p^{th}$ = 12.5 deg., the excitation takes place only when the parametric resonance condition is fulfilled exactly. For larger  $\theta_p$, where PEMs are excited over a wider field interval, a larger frequency mismatch is allowed (for details see Eq. (2) in [\onlinecite{Urazhdin2010}]). It is also clear that the parametric process leads to mode hybridization which results in both a slight reduction of the droplet frequency and to the formation of higher harmonics.

\noindent\textbf{\uppercase\expandafter{\romannumeral4}. EXPERIMENTAL DEVICE FABRICATION}

\begin{figure}[h]
\centering
\includegraphics[width=0.48\textwidth]{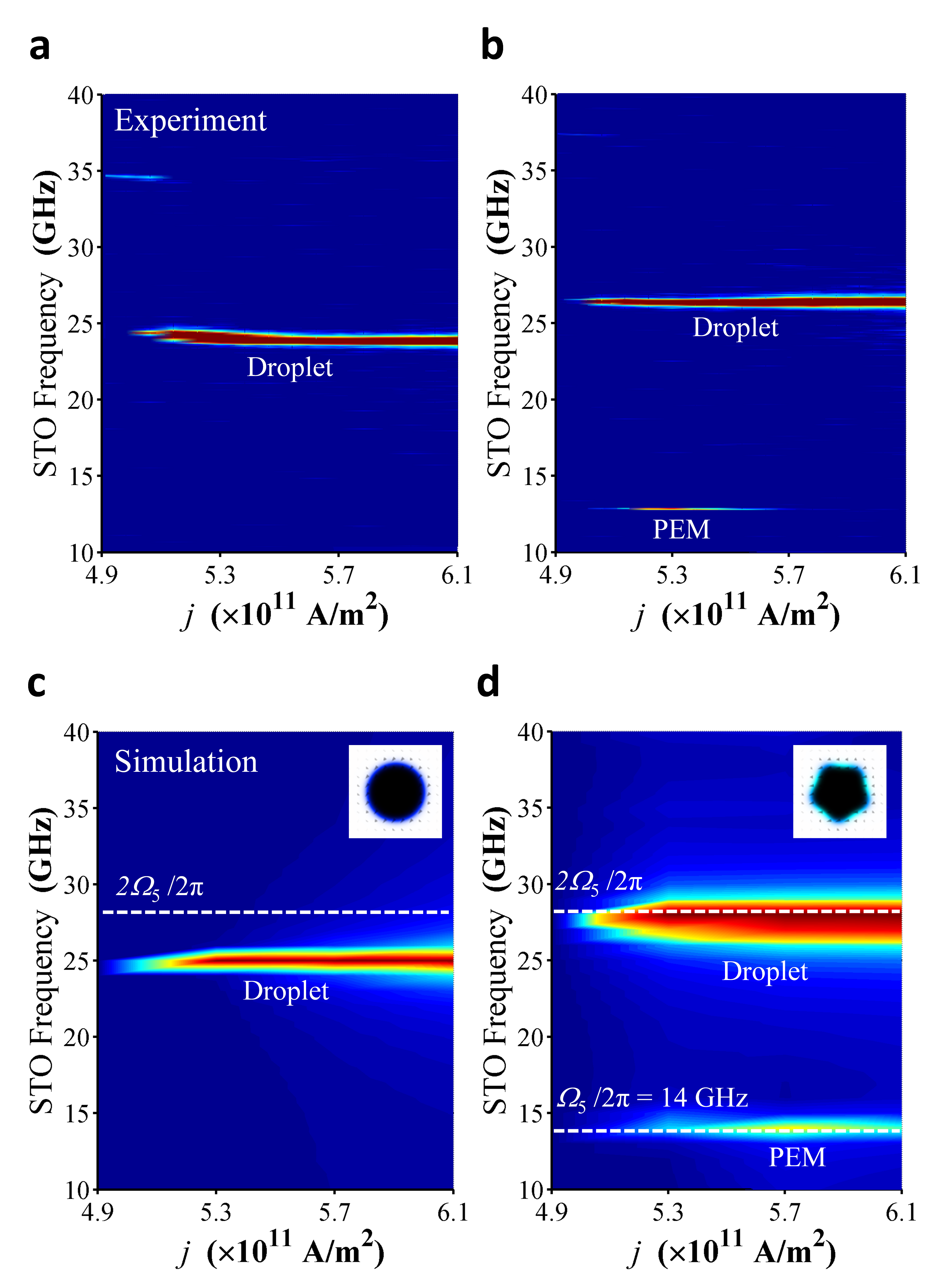}
\caption{\noindent{Comparison of micromagnetic simulations with experimental data on droplet PEM.} Experimental measured phase diagram of NC-STO on Co/Cu/Co-[Ni/Co]$_{\times4}$ orthogonal spin-valve with an applied magnetic field of (a) 0.8 T and (b) 0.9 T with a tilt angle of 7.5 deg. Simulated phase diagram of the same device with slightly different field strengths and tilt angle as compared to the experimental data: (c) $H$ = 0.7 T, (d) $H$ = 0.8 T at a tilt angle of 3 deg. Other material parameters used in the simulation include: $R_c$ = 60 nm, $A_{ex}$ = 30 pJ/m, $K$ = 447 kJ/m3 and $M_s$ = 716 kA/m.}
\label{Fig:comparison}
\end{figure}

We finally turn to experiments. Fig.\ref{Fig:comparison} shows the experimental and calculated microwave power of a nanocontact spin torque oscillator (NC-STO) based on a Co/Cu/Co-[Ni/Co]$_{\times4}$ orthogonal spin-valve in different applied magnetic field strengths and tilt angles. In the experiment, the PEM is excited in a field of 0.9 T tilted 7.5$^\circ$ from the film normal. This is reasonably close to the simulations which use a field of 0.8 T tilted 3$^\circ$, with all material parameters being similar to experimentally measured values \cite{Mohseni2013}. From Eqs. (9-12), the PEM frequency of 14 GHz %in the phase diagram
corresponds to mode number $n$ = 5, which is indeed observed in our numerical simulations as shown in the inset of Fig.\ref{Fig:comparison}(d). %It is worth noting that all the material parameters used in the micromagnetics simulations are the same with the experimental values \cite{Mohseni2013}.
Therefore, our theoretical model presented in this work is able to correctly identify the  different dynamical states. Moreover, it is possible to theoretically predict the different parameter spaces for exciting or inhibiting these PEMs, which will be an important design feature in %states, including droplet precession, breathing and PEM \emph{etc} for
real applications.

Orthogonal pseudo-spin-valve stacks consisting of Co (6 nm) / Cu (6 nm) / Co (0.2 nm)[Ni (0.6 nm)/Co (0.25 nm)]$_{\times4}$ with a Ta (4 nm) /Cu (10 nm) / Ta (4 nm) seed layer and a Cu (2 nm) / Pd (2 nm) cap layer were magnetron sputter deposited on thermally oxidized Si wafers. The stack was patterned into 8 $\times$ 16 $\upmu$m$^2$ mesas using optical lithography, and then coated with a 30 nm $\mathrm{SiO_{2}}$ interlayer dielectric deposited by chemical vapor deposition. NCs were fabricated using electron beam-lithography and reactive ion etching through the SiO$_2$. Finally, 1.1 $\upmu$m Cu top contacts in the shape of microwave wave guides were fabricated by optical lithography, sputter deposition, and lift-off. The microwave signal properties of the devices were characterized in our custom probe station where fields up to 2 T can be applied, while the field angle can be varied by tilting the sample using mechanical rotation. The device under test is driven by a direct current provided by a Keithley 6221 current source and the device resistance is monitored using a Keithley 2182A nanovoltmeter. The generated microwave voltage is decoupled from the dc bias using a 40 GHz bias-T. The microwave signal is then amplified with a 0.1-30 GHz low noise +26 dB amplifier and analyzed in the frequency domain using a R\&S FSU spectrum analyzer.
\\

\noindent\textbf{\uppercase\expandafter{\romannumeral5}. CONCLUSION}

We have studied perimeter excitations of magnon drops and magnetic droplet solitons. We have shown that there exists a family of such excitations that can be described as standing waves of periodic deformation of the droplet perimeter. The frequency of these modes is described by a simple equation [Eq. (\ref{Omegans})] and the damping rate for sufficiently small mode indices follows a Gilbert relation with an effective damping constant $\alpha_p$ several times larger than the Gilbert constant of the magnetic material. We have also demonstrated that the perimeter may become unstable in a droplet driven by a sufficiently large spin-polarized current if the current polarization direction is inclined with respect to the film normal. This instability has a parametric character and has the minimum threshold when the exact condition of a parametric resonance is fulfilled. These findings not only provide new information about the structure and properties of internal excitations of droplets, but are also important for the practical use of spin-torque nano-oscillators based on current-driven droplets for microwave frequency signal generation and processing.
\\\\

\noindent\textbf{Acknowledgements}

Y.Z. acknowledges support by the National Natural Science Foundation of China (Project No. 1157040329) and Shenzhen Fundamental Research Fund under Grant No. JCYJ20160331164412545. Y. W. Liu thanks for the support by the National Basic Research Program of China (Grant No. 2015CB921501) and NSFC (Grant No. 11274241 and No. 51471118).  A.S. and V.T.  were supported by  the Grant ECCS-1305586 from the National Science Foundation of the USA, by the contract from the US Army TARDEC, RDECOM, by the DARPA grant ``Coherent Information Transduction between Photons, Magnons, and Electric Charge Carriers'' and by the Center for NanoFerroic Devices (CNFD) and the Nanoelectronics Research Initiative (NRI). J.\AA. was supported by the ERC Starting Grant 307144 "MUSTANG", the Swedish Foundation for Strategic Research (SSF) (Successful Research Leaders), the Swedish Research Council (VR), and the Knut and Alice Wallenberg Foundation.
\\\\
\noindent\textbf{Author contributions}

Y. Z. conceived and designed the project. Xiao Dun, Y. H. Liu, Y. W. Liu and Y. Z. carried out the numerical simulations. V. T. and A. S. performed the theoretical analysis. S. M. M., S. C. and J.\AA. contributed in device fabrication and carried out all measurements. All authors interpreted the data and contributed to preparing the manuscript. Correspondence and requests for materials should be addressed to Y. Z.
\\\\
\bibliographystyle{apsrev4-1}

\end{document}